\documentclass[
reprint,
superscriptaddress,
 amsmath,amssymb,
 aps,
prstab,
floatfix,
]{revtex4-1}

\usepackage{graphicx}
\usepackage{dcolumn}
\usepackage{bm}
\usepackage[english]{babel}
\usepackage[utf8]{inputenc}
\usepackage{siunitx}
\usepackage{blindtext}
\usepackage{physics}
\usepackage{mathtools} 

\usepackage[dvipsnames]{xcolor}
\usepackage{hyperref}
\hypersetup{
    colorlinks=true,
    linkcolor=blue,
    filecolor=magenta,      
    urlcolor=cyan,
}
\usepackage{cleveref}
\makeatletter
\renewcommand{\fnum@figure}{\normalfont FIG. \thefigure}
\renewcommand*{\@caption@fignum@sep}{\normalfont . }

\newcommand{\jeti}{{JETi200}~}

\makeatother

\begin{document}

\preprint{APS/123-QED}

\title{Decoding the electron bunch duration via passive plasma lensing}
\title{Determination of few femtosecond to attosecond electron bunch durations using a passive plasma lens}
\author{Andreas Seidel}
\email{seidel.andreas@uni-jena.de}
\affiliation{Friedrich-Schiller-Universit{\"a}t, F{\"u}rstengraben 1, 07743 Jena, Germany}
\affiliation{Helmholtz-Institut Jena, Fröbelstieg 3, 07743 Jena, Germany}
\affiliation {GSI GmbH, Planckstr 1, 64291 Darmstadt, Germany}
\author{Carola Zepter}
\affiliation{Friedrich-Schiller-Universit{\"a}t, F{\"u}rstengraben 1, 07743 Jena, Germany}
\affiliation{Helmholtz-Institut Jena, Fröbelstieg 3, 07743 Jena, Germany}
\author{Alexander S{\"a}vert}

\affiliation{Friedrich-Schiller-Universit{\"a}t, F{\"u}rstengraben 1, 07743 Jena, Germany}
\affiliation{Helmholtz-Institut Jena, Fröbelstieg 3, 07743 Jena, Germany}
\affiliation {GSI GmbH, Planckstr 1, 64291 Darmstadt, Germany}

\author{Stephan Kuschel}
\email{stephan.kuschel@tu-darmstadt.de}
\affiliation{Technische Universit{\"a}t Darmstadt, Karolinenplatz 5, 64289 Darmstadt, Germany}
\author{Matt Zepf}
\affiliation{Friedrich-Schiller-Universit{\"a}t, F{\"u}rstengraben 1, 07743 Jena, Germany}
\affiliation{Helmholtz-Institut Jena, Fröbelstieg 3, 07743 Jena, Germany}
\affiliation {GSI GmbH, Planckstr 1, 64291 Darmstadt, Germany}
\date{\today}

\begin{abstract}
Determining the pulse duration of femtosecond electron bunches is challenging and often experimentally invasive. An effective method for measuring the duration based on the time-dependent variations in electron beam divergence induced by a passive plasma lens is described. 

Reconstruction of the temporal shape of the electron bunch down to $c \cdot dt=\SI{10}{\nano\meter}$ ($\sim \SI{30}{\atto\second}$)  without external RF-cavities or multi-octave spanning spectrometer is shown numerically. Experimental data from a $\sim$ 3fs electron bunch demonstrates practical applicability of this method. While this approach can be used with any high current electron beam, it is particularly well matched to laser-driven and particle-driven wakefield accelerators and also accommodates electron beams with a time-dependent beam-energy (eg. 'chirp').
\end{abstract}

\maketitle


\section{Introduction}
Laser wakefield acceleration (LWFA)\cite{tajima_laser_1979} has emerged as a technique capable of generating GeV-class electron beams \cite{gonsalves_petawatt_2019} with femtosecond-scale durations \cite{Zarini_BunchLength} and kiloampere-level peak currents \cite{huang_electro-optic_2024}. These ultrashort, high-charge-density bunches enable novel applications ranging from compact free-electron lasers \cite{galletti_prospects_2024}, hybrid acceleration \cite{hidding2023} towards future high-energy physics experiments \cite{foster_hybrid_2023}, or use LWFA as an injector to replenish a synchrotron \cite{antipov2021}.

Despite rapid progress, resolving the temporal structure of ultrafast electron bunches, particularly for durations below \(\tau<\SI{10}{\femto\second}\), still remains a challenge. 
Typical accelerator diagnostics, such as X-band radio-frequency cavities \cite{Behrens2014, Dolgashev_2014_x-band} face resolution limits at a sub \SI{10}{\femto\second} \cite{maxson_direct_2017} and require temporal synchronization. Faraday rotation probes\cite{Buck_2011_ProbingFaradayElectronbunch} struggle with signal-to-noise ratios at sub-10 fs scales. Multi-octave spectrometers can provide a measurement of the temporal substructure of the electron bunch on the scale from 0.7 to 40\,fs\cite{Zarini_BunchLength}, but they are challenging to setup, operate and require iterative Fourier-based reconstruction algorithms, which deduce the bunch shape from the reconstructed phase.

Plasma lenses are a distinct and new class of electron bunch diagnostic. Active plasma lenses, for example, have been used to measure  the emittance. 
Unlike magnetic lenses, active plasma lenses provide kT/m-scale focusing gradients while simultaneously acting as temporal diagnostic tools. Their chromatic properties enable emittance characterization \cite{Barber2017} down to a $\mu m$-scale level. However, alignment sensitivity has a crucial impact on the measurement \cite{Lindstroem_2018_emittance}. 

Here, we discuss the use of passive plasma lenses as a pulse duration diagnostic. Passive plasma lensing offers a disruptive alternative to determining electron bunch parameters by exploiting inherent beam-plasma interactions without the need for fine alignment. The extremely rapid response of a passive plasma lens at sufficiently high density offers excellent temporal resolution. In vacuum, a relativistic electron bunch creates magnetic ($B_\theta$) and electric ($E_r$) fields. The resulting focusing ($cB_\theta$) and defocusing ($E_r$) forces cancel in the ultra-relativistic limit ($\gamma \rightarrow\infty$), such that $E_r-cB_\theta \rightarrow 0$. 

When a relativistic electron bunch passes through a plasma, two distinct regimes govern the dynamics: the overdense regime (\( n_p > n_b \)), where the plasma density (\( n_p \)) exceeds the density of the electron bunch (\( n_b \)), and the underdense regime (\( n_p < n_b \)). Self-focusing is possible in both regimes.
In the overdense regime, the background plasma electrons neutralize the electric field partially, resulting in the focusing of the electron bunch driven by the remaining self-generated magnetic field. Conversely, in the underdense regime, the electron bunch significantly modulates the background plasma density, resulting in strong transverse and longitudinal electromagnetic fields, which in turn generate intense focusing forces on the electron bunch \cite{Kuschel_passiveLens,Thaury2015,doss_laser-ionized_2019,doss_underdense_2023}.

The rapid response of the plasma lens to the temporally varying bunch parameters (particle density, energy chirp) result in spectrally dependent divergence changes. This allows the temporal shape of the electron bunch to be reconstructed with sub-femtosecond accuracy without the need for temporal synchronization or precise alignment.

\section{Theoretical Model}

\subsection{Wakefield generation and focusing fields}

Transverse wakefields generated by relativistic electron bunches propagating through plasma are analyzed using linear wakefield theory \cite{Keinigs_1987_PlasmawaveTheory,Vera_2022_wakefield-theory}. The electron bunch density distribution can be represented as \(n_b(\xi=ct-z,r)=n_{b0}\cdot n_{b||}(\xi)\cdot n_{b\perp}(r)\) where $n_{b0}$ is the peak charge density, while $n_{b||}$ and $n_{b\perp}$ are the normalized longitudinal and radial charge distributions, respectively. The transverse wakefield, defined as \(W_\perp=E_r-cB_\theta\) can be expressed analytically as \cite[eqn. 63]{Keinigs_1987_PlasmawaveTheory}: 

\begin{equation}
    \label{eq: Wperp}
        W_\perp(\xi,r)=\frac{-n_{b0}e}{\epsilon_0k_{pe}}\dv{R(r)}{r}\int_{-\infty}^\xi n_{b||}(\xi')\cdot\sin{\left(k_{pe}(\xi-\xi')\right)}\,d \xi'
\end{equation}
Where $R(r)$ is given by: 
\begin{equation*}
\begin{split}
        R(r)&=k_{pe}^2K_0(k_{pe}r)\int_0^rr'n_{b\perp}(r')I_0(k_{pe}r')dr'\\
        &+k_{pe}^2I_0(k_{pe}r)\int_r^\infty r'n_{b\perp}(r')K_0(k_{pe}r')dr'
\end{split}
\end{equation*}
With $K_0$ and $I_0$ representing modified Bessel functions and $k_{pe}$ the plasma wave number. To prevent beam filamentation, the threshold condition $k_{pe}\cdot n_{b\perp}\leq 2.2$ has been found \cite{Allen_2012_fillamentation}.

\begin{figure}
    \centering
    \includegraphics[width=1\linewidth]{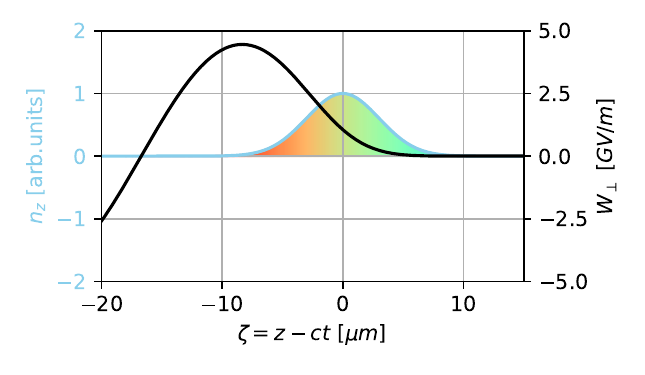}
    \caption{Analytical transverse wakefield W$_\perp$ from \cref{eq: Wperp} for a chirped electron bunch with rms beam length $\sigma_z = \SI{3}{\micro\meter}$ and rms beamsize $\sigma_r = \SI{6.8}{\micro\meter}$ driving a linear wakefield in a plasma with $n_p=\SI{1e18}{\per\centi\meter\cubed}$. The total charge of the electron beam is \(Q=\SI{65}{\pico\coulomb}\). In light blue, the normalized longitudinal charge distribution, and in black, the focusing fields} 
    \label{fig:analyticalResults}
\end{figure}

The transverse wakefield generated by an electron bunch with $\sigma_z=\SI{3}{\micro\meter}$ is shown in figure \ref{fig:analyticalResults}. We assume a chirped electron bunch, as indicated by the color scale. The head of the bunch contains the most energetic electrons while its tail contains the least energetic electrons. Due to the self-induced transverse wakefield, the head of the bunch (high energy electrons) experiences weaker focusing fields than the trailing part (low energy electrons). Consequently this leads to a strong variation of the electron focusing over the length of the bunch,  
resulting in very different transverse momenta easily measured using  magnetic electron spectrometers with angular resolution.

\begin{figure}[b]
    \centering
    \includegraphics[width=1\linewidth]{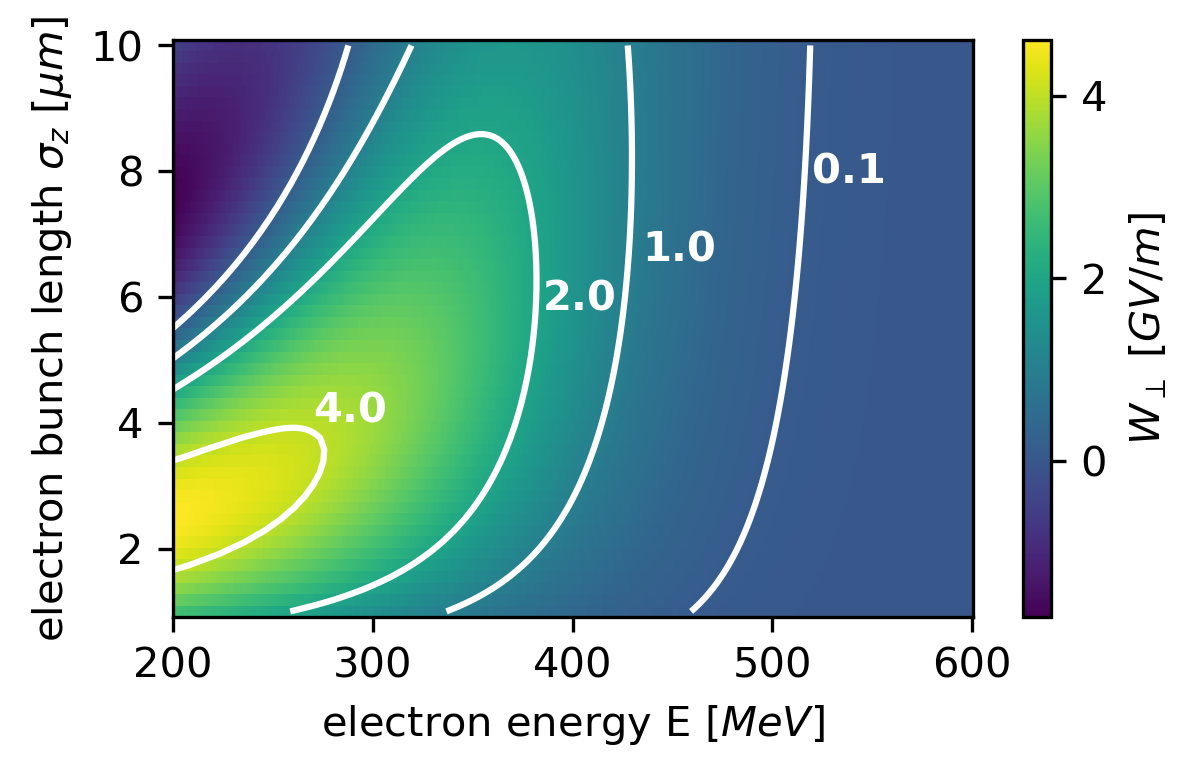}
    \caption{Analytical result of focusing fields W$_\perp$ from \cref{eq: Wperp} for different electron bunch lengths. The total charge of the Gaussian electron beam was kept constant at \(Q=\SI{65}{\pico\coulomb}\).} 
    \label{fig:analyticalResultsOverEnergy}
\end{figure}

This dependence is shown in Figure \ref{fig:analyticalResultsOverEnergy}: the transverse wakefield strength varies significantly with both electron energy and bunch length. While maintaining a constant charge of \SI{65}{\pico\coulomb} within a plasma density of $n_p=\SI{1e18}{\per\centi\meter\cubed}$, we observe that electrons exceeding \SI{500}{\mega\electronvolt} experience almost no focusing regardless of bunch length. However, at lower energies (\SI{250}{\mega\electronvolt}), the transverse field strength peaks at a bunch length of $\sigma_z=\SI{3}{\micro\meter}$, becoming weaker for both shorter and longer bunches. These distinctive focusing patterns create a unique energy-dependent divergence change in the electron bunch, which enables the reconstruction of the longitudinal bunch profile. 
The analytical description is valid, when the induced wakefield is in the linear regime, the electron bunch dimensions are smaller than one plasma wavelength, and the electron bunch is monotonically chirped. Its important to note that the actual dependence $E(\zeta)$ can be modeled and reconstructed with our method as long as it is monotonically increasing.

As suggested by the the separatrix model \cite{Esarey_1995_separatrix}, earlier-injected electrons achieve higher energies before reaching dephasing than those injected later, resulting in a broad energy spectrum for bunches generated via ionization injection compared to localized injection mechanisms. Additional positive chirp contributions arise from path length differences during drift—for example, \SI{700}{\mega\electronvolt} and \SI{100}{\mega\electronvolt} electrons traveling \SI{100}{\nano\meter} develop a path difference of \SI{1.3}{\micro\meter}, further enhancing the chirp and the observable effects in the energy spectrum.

To quantitatively model the focusing strength experienced by each electron energy inside the electron bunch, we represent the longitudinal electron bunch profile as a sum of discrete, Gaussian-shaped bunchlets, $n_i(\xi_i)$, each with an rms length of 100\,nm. These bunchlets contain electrons within a \SI{1}{\mega\electronvolt} energy interval and are located at longitudinal positions, $x_i$. 
The total longitudinal profile is then given by \(n_{b||}(\xi)=\sum_{i=E_1}^{E_n}n_{i}(\xi+\xi_i)\). This electron density distribution generates a wakefield, which in turn produces transverse focusing fields that can be calculated using \cref{eq: Wperp}.\\

\begin{figure}[t]
    \centering
    \includegraphics[width=1\linewidth]{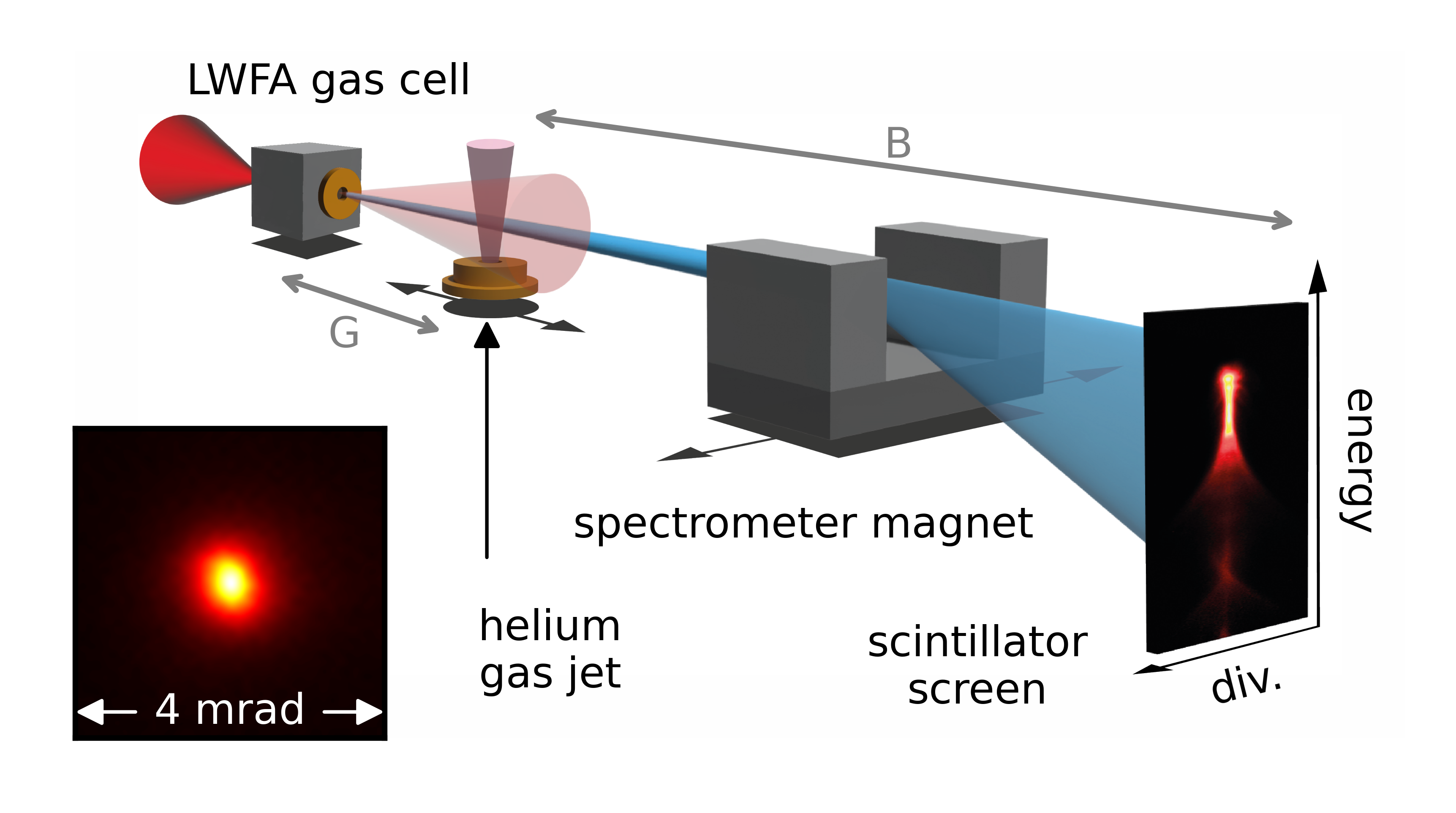}
    \caption{Schematic of the experimental setup. The \jeti laser is focused with an $f/24$ parabolic mirror onto a gas cell. The distance between the gas cell and the consecutive jet acting as plasma lens could be varied. A movable dipole magnet disperses the electrons on the scintillation screen, enabling measurement of the energy spectrum and transverse beam profile. Left inset: Transverse beam profile}
    \label{fig:setup}
\end{figure}
The focusing strength for each bunchlet is characterized by the parameter $\kappa$.
\begin{equation}
    \label{eq:kappa_gamma}
    \kappa(\gamma,\xi)=\frac{e}{p_e}\pdv{W_\perp(\xi)}{r}
\end{equation}
where $p_e=p_e(\gamma)$ is the momentum of the bunchlet and $\pdv{W_\perp(\xi)}{r}$
is the radial derivative of the transverse focusing field. Since each bunchlet $n_i(\xi_i)$ is located at a different longitudinal position, it experiences a distinct focusing wakefield and thus a unique focusing strength parameter $\kappa(\gamma,\xi)$.

\subsection{transverse beam dynamics}
The transverse dynamics of relativistic charged particles can be described using the transfer matrix formalism. For a single particle with initial phase-space coordinates ($x_0$,$x_0'$), propagation through an optical system transforms the coordinates:  
\begin{equation*}
\begin{pmatrix}
        x\\
        x'
    \end{pmatrix}=\tilde{M}\begin{pmatrix}
        x_0\\
        x_0'
\end{pmatrix}
\end{equation*}
where $\tilde{M}$ represents the cumulative transfer matrix formed by multiplying individual component matrices ($M_\text{drift}$,$M_\text{lens}$,etc.) \cite{Reiser_2008_ElectronBasic}. In a typical quadrupole lens the result will have the form of eq.\ref{eq:quadrupolMatrix}. 
\begin{widetext}
\begin{equation}
\begingroup 
\setlength\arraycolsep{10pt}
\tilde{M}=
    \begin{pmatrix*}[r]
    \cos(\sqrt{\kappa}L) - B\sqrt{\kappa}\sin(\sqrt{\kappa }L)& G\cos(\sqrt{\kappa}L)+\frac{1}{\sqrt{\kappa}}\sin(\sqrt{\kappa }L)+B(- G\sqrt{\kappa}\sin(\sqrt{\kappa }L)+\cos(\sqrt{\kappa}L))\\
    -\sqrt{\kappa}\sin(\sqrt{\kappa }L)&- G\sqrt{\kappa}\sin(\sqrt{\kappa }L)+\cos(\sqrt{\kappa}L)
\end{pmatrix*}
\label{eq:quadrupolMatrix}
\endgroup
\end{equation}
\end{widetext}

Here, G and B denote drift lengths before/after the lens, L the quadrupole length, and $\kappa=\frac{1}{Lf_0\gamma/\gamma_0}$ the normalized focusing strength, where $f_0$ is the reference focal length at energy $\gamma_0 mc^2$. For our plasma lens, this is replaced by \cref{eq:kappa_gamma}.

For a \textit{particle ensemble}, this formalism extends to beam size and divergence via the covariance matrix $\Sigma = \langle \mathbf{X} \mathbf{X}^\top \rangle$, where $\mathbf{X} = (\sigma_0, \sigma_0')^\top$. Under the Courant-Snyder parameterization ($\alpha$, $\beta$, $\gamma$) \cite{Courant_1958_courant-snyder-paramter}, the geometric emittance is conserved as $\epsilon = \sqrt{\det \Sigma}$. Assuming a beam waist ($\alpha = 0$) at the plasma source exit, the final rms beam size at the spectrometer reduces to:
\begin{equation}
    \label{eq:sigmaf}
    \sigma_f = \sqrt{\tilde{M}_{11}^2 \sigma_0^2 + \tilde{M}_{12}^2 {\sigma'_0}^2},
\end{equation}

where $\sigma_0$ and $\sigma_0'$ are the initial rms size and divergence. We use this formula to calculate the beam size on the spectrometer screen and the parameter $\kappa$ is given by the wakefield, eq.~(\ref{eq:kappa_gamma}).

\section{Experiment}
\subsection{Setup}
\begin{figure*}[t]
    \includegraphics[width=1\linewidth]{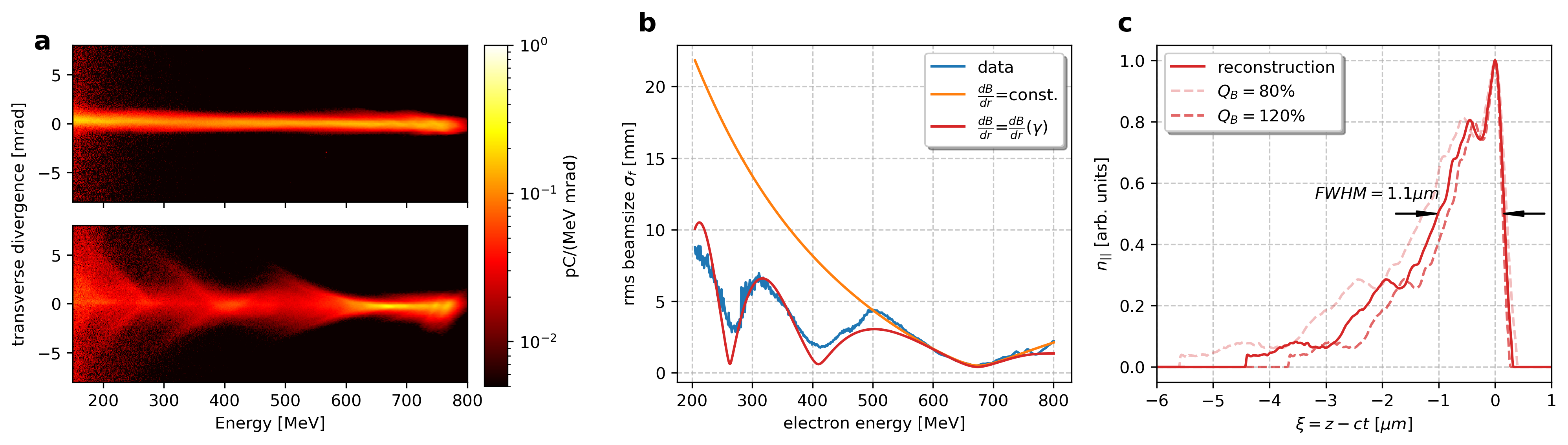}
    \caption{(a) Observed electron beam energy spectrum without (a, top) and with (a, bottom) plasma lens, showing the transverse divergence as a function of energy. Pronounced modulations in the beam divergence are observed across the spectrum with a plasma lens. (b) Root-mean-square (rms) beam size as a function of electron energy. The blue curve represents experimental data extracted from vertical lineouts of the spectrum in (a2). The orange and red curves correspond to results from different fitting algorithms, including constant dB/dr (orange), energy-dependent dB/dr $(\gamma)$ (red). (c) Reconstruction of the longitudinal electron bunch profile corresponding to the analytic model underlying the result in (b) (red curve). Dashed curves indicate the solutions obtained assuming different  bunch charges with \SI{80}{\percent} and \SI{120}{\percent} of nominal bunch charge shown respectively.}
    \label{fig:experimental_results}
\end{figure*}
An experiment was conducted which allowed the temporal response of the passive plasma lens to be tested. The experiment was conducted at the \jeti laser system at the Helmholtz Institute Jena (set-up shown in \cref{fig:setup}). The laser system delivered pulses with energies up to $E =\SI{4.0}{\joule}$ at a central wavelength of $\lambda_0=\SI{800}{\nano\meter}$ and duration of $\tau=\SI{23}{\femto\second}$. 
An f/24 off-axis parabolic mirror focused the beam to a vacuum full-width at half-maximum (FWHM) focal spot diameter of \SI{22}{\micro\meter}, with approximately \SI{40}{\percent} of the total energy contained within the FWHM, resulting in a normalized vector potential of \( a_0 = 2.9 \).

The laser beam was focused into the entrance of a \SI{7.0}{\milli\meter}-long gas cell, developed at the Helmholtz Institute Jena. It was filled with a \SI{92}{\percent}-helium and \SI{8}{\percent}-nitrogen mixture by volume. The higher nitrogen concentration compared to the commonly used 1 to \SI{5}{\percent} used for ionization injection, maximized electron beam charge while maintaining adequate laser guiding. ANSYS\cite{ANSYSFluentGuide2019} fluid dynamics simulations of the gas flow through the cell revealed a central 6-mm-long uniform density region with 1.5\,mm long density ramps on either side.

The secondary target, used for plasma lensing, was a \SI{7}{\milli\meter} wide Laval nozzle operated with pure helium, positioned at a variable distance between \SI{4}{\milli\meter} and \SI{125}{\milli\meter} behind the gas cell exit. Interferometric measurements\cite{Hüther_gasjet_density} confirmed that the neutral gas density profile above the nozzle remained stable within \SI{5}{\percent} at a height of \SI{2}{\milli\meter} above the nozzle.

A Kodak BioMAX scintillation screen\cite{Kurz_2018_Lanex} positioned \SI{237}{\centi\meter} downstream from the gas cell recorded the electron beam profiles and spectra. A \SI{10}{\centi\meter} long dipole magnet could be inserted into the beam path, allowing for measurement of either the electron beam profile or energy spectrum on the same screen. The optical system was calibrated using a \SI{530}{\nano\meter} laser diode in conjunction with established scintillator response curves\cite{Kurz_2018_Lanex}, enabling absolute charge measurements.

\subsection{Passive plasma lens results}
The experimental results of the electron beam energy spectrum after acceleration in the gas cell and subsequent propagation through a plasma lens are presented in Figure \ref{fig:experimental_results}a). In this configuration, the distance between the acceleration stage and the plasma lens (second gas jet) was $G = \SI{7}{\milli\meter}$, with the plasma lens itself being \SI{7}{\milli\meter} long and the spectrometer screen positioned $B=\SI{2360}{\milli\meter}$ from the plasma lens. The data shows clear focusing of the electron beam at energies of \SI{650}{\mega\electronvolt} and \SI{415}{\mega\electronvolt}, while significant defocusing is observed at \SI{500}{\mega\electronvolt} and \SI{310}{\mega\electronvolt}, see Figure \ref{fig:experimental_results}a, bottom. Notably, the highest electron energies (E$>$\SI{750}{\mega\electronvolt}) remain largely unaffected by the plasma lens, as predicted by the theory. 

These focusing-defocusing features are more clearly illustrated in Figure \ref{fig:experimental_results}b, where the rms beam size on the screen as a function of electron energy is shown in blue. This panel also compares the experimental data to several theoretical models. Given the precisely known distances in our setup, we apply \cref{eq:quadrupolMatrix,eq:sigmaf} under the assumption of a constant magnetic field gradient, which is typical for magnetic quadrupole lenses or solenoids, but clearly violated for our plasma lens. The resulting beam size (orange curve) accurately describes the high-energy portion of the spectrum ($E >$ \SI{500}{\mega\electronvolt}), but fails to capture the pronounced oscillations in beam divergence observed at lower energies. Such oscillatory behavior is  predicted at much lower electron energies ($E <$ \SI{50}{\mega\electronvolt}) for a constant-gradient lens (see Appendix \ref{appendix:C}). This observation agrees with the prediction, based on wakefield theory, that the electron bunch is positively chirped, meaning that lower-energy electrons are positioned at the rear of the bunch and thus experience stronger focusing fields.

Further including \cref{eq:kappa_gamma} into the calculation and using the model described earlier, the final beam size on the spectrometer is shown in fig. \ref{fig:experimental_results}b (red curve). We iteratively adjusted the positions  $\xi_i$ of the bunchlets and recalculated the resulting focusing field $W_\perp(\xi)$ and final beam size to minimize the differences between the modeled and the measured transverse beam sizes on the spectrometer. Details of this optimization procedure are provided in the supplementary material.

The result of this optimization procedure (Figure \ref{fig:experimental_results}b, red curve) exhibits excellent agreement with the experimental data, accurately reproducing both the positions of the minima and maxima in the rms beam size across the entire energy range. 
The final reconstructed longitudinal bunch profile $n_{b||}(\xi)$, based on the optimized positions $\xi_i$ of all bunchlets, is shown in Figure \ref{fig:experimental_results}c. The dashed lines show the reconstructions with $\pm\SI{20}{\percent}$ bunch charge. The reconstructed bunch length is consistent with previous studies\cite{Zarini_BunchLength,Buck_2011_ProbingFaradayElectronbunch}. \\

\subsection{Comparison PIC-Simulation and theoretical reconstruction}
\begin{figure}[b]
    \includegraphics[width=0.8\linewidth]{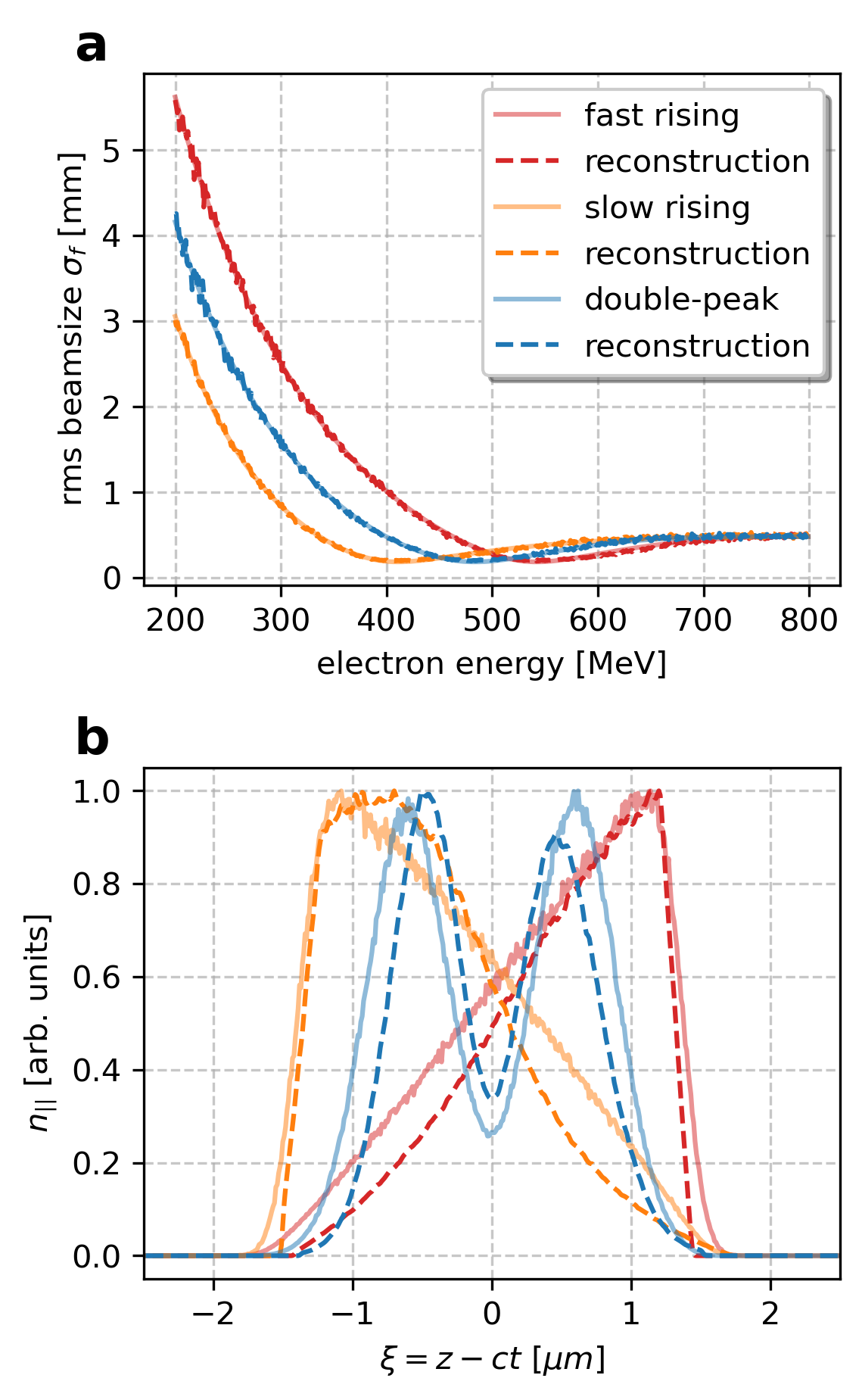}
    \caption{(a) Root-mean-square (rms) beam size as a function of electron energy for electron bunchs with different longitudinal profile after plasma lensing, comparing PIC simulation results (solid) and analytical modeling (dashed) for optimized lens parameters (rectangular density profile, $n_P=$\SI{1e17}{\per\cubic\centi\meter} and $L=\SI{2}{\milli\meter}$). (b) Comparison of the reconstructed longitudinal bunch profiles from analytic modeling (dashed) and the profile used as input for the PIC simulation (solid), 
   showing closely matched pulse shape 
   and overall bunch length.}
    \label{fig:PIC-vs-Analytic}
\end{figure}
We performed particle-in-cell (PIC) simulations using FBPIC \cite{lehe_spectral_2016} to fully model the passive plasma lens interaction. The simulation domain in longitudinal and radial direction was set to \SI{50}{\micro\meter} by \SI{20}{\micro\meter} (3200 x 200 gridpoints) using $N_m=2$ azimuthal modes. 

We simulated an electron bunch with an intitial rms divergence of \SI{0.2}{\milli\radian}

while retaining key parameters of the experiment: (\SI{75}{\pico\coulomb}) charge, a triangular energy spectrum (200-\SI{800}{\mega\electronvolt}, peaking at \SI{750}{\mega\electronvolt}), rms source size (\SI{1}{\micro\meter}), and total bunch length (\SI{3}{\micro\meter}). The electron source-to-plasma lens distance was increased to \SI{40}{\milli\meter} (from \SI{7}{\milli\meter} in the experiment (resulting in the same charge density at the plasma lens) and maintaining a total source-to-spectrometer distance of \SI{2370}{\milli\meter}. The lens was modeled with a rectangular density profile ($n_P=\SI{1e17}{\per\cubic\centi\meter}$ and $L=\SI{2}{\milli\meter}$), deliberately reducing the focusing field strength $W_\perp$ to suppress multi-energy oscillations to test our focusing model.

The parameters chosen for the simulation exhibit a single focus as shown in figure \ref{fig:PIC-vs-Analytic}a. We find that this improves the fidelity of reconstruction. 
Applying the analytic model from Figure \ref{fig:experimental_results}b (red curve) to this configuration yields excellent agreement with the simulation (dashed lines), confirming the model’s validity under controlled conditions. The reconstructed longitudinal bunch profile from the analytic reconstruction (Figure \ref{fig:PIC-vs-Analytic}b dashed lines) align closely with the input PIC simulation profiles (solid curves), matching both the total bunch length (\SI{3}{\micro\meter}) and the overall shape. This correspondence validates the analytic method’s ability to reconstruct longitudinal bunch structures with high fidelity when phase-space perturbations are minimized. The rising (falling) edge of the pulses in figure \ref{fig:PIC-vs-Analytic} have a 10-90$\%$ edge of \SI{1}{\femto\second} ($\sim \SI{300}{\nano\meter}$) showing the high temporal resolution obtainable.

\begin{figure}[b]
    \includegraphics[width=0.95\linewidth]{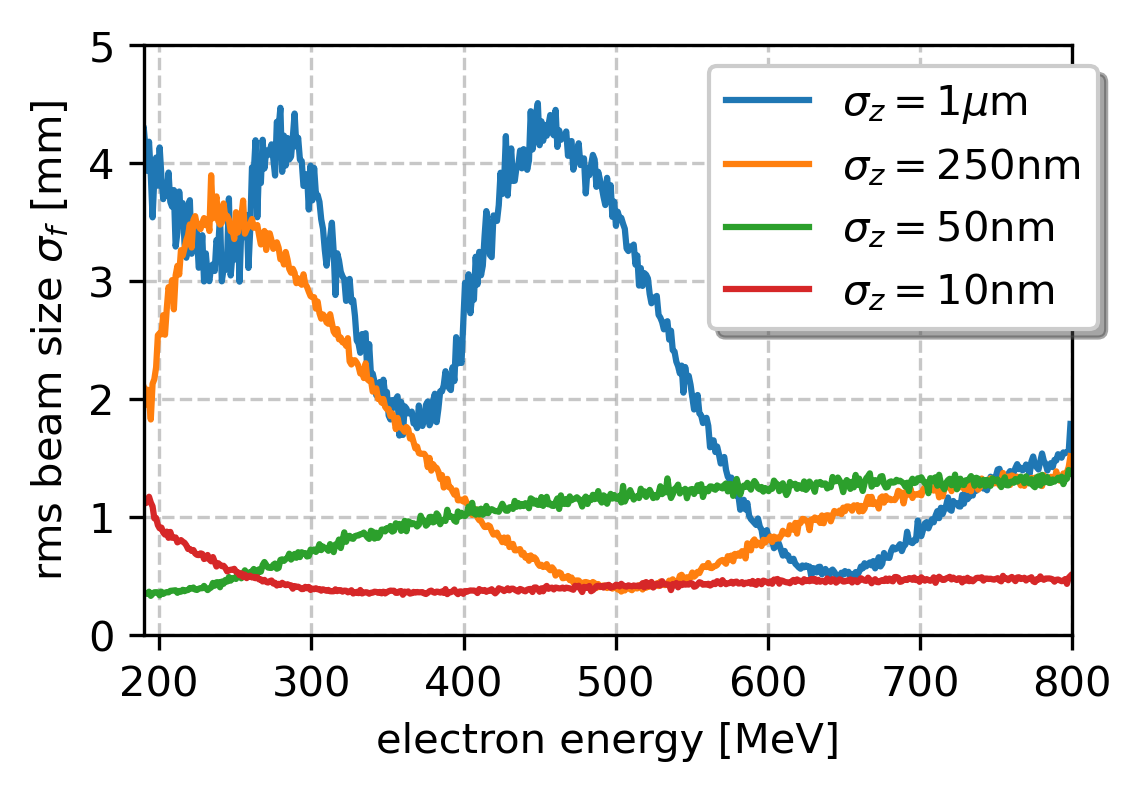}
    \caption{PIC-simulation results of the rms beam size versus electron energy for total bunch lengths ($2\sigma_z$) from \SI{2}{\micro\meter} to \SI{20}{\nano\meter}. The electron bunch charge density is constant for all cases, while the plasma lens density is scaled as $n_P$ = \SI{8.5e17}{\per\cubic\centi\meter} $\cdot \frac{\SI{1}{\micro\meter}}{\sigma_z}$. Shorter bunches exhibit reduced beam size oscillations, due to weaker focusing fields $W_\perp$. The plasma lens profile was changed from gaussian to rectangular, and the divergence was reduced to \SI{0.2}{\milli\radian} for the shortest bunch $\sigma_z$=\SI{10}{\nano\meter}  to increase focusing while adhering to \cref{eq:plasmalimit}}
    \label{fig:PIC-beamlength}
\end{figure}

We also performed simulations comparing the transverse beam size dynamics for electron bunches ranging from \SI{2}{\micro\meter} to \SI{100}{\nano\meter} (total bunch length, not rms) in order to systematically quantify the impact of bunch length on the electron beam focusing in figure \ref{fig:PIC-beamlength}. All configurations retained the experimental longitudinal plasma density profile and constant electron bunch charge density. For short bunches ($k_p\sigma_z\ll 1$, where $\sigma_z<\SI{3}{\micro\meter}$; see Figure \ref{fig:analyticalResultsOverEnergy}), wakefield theory predicts weaker transverse focusing fields at the same electron energy due to reduced overlap with the plasma wake. To counteract this, we scaled the plasma lens density inversely with bunch length: ($n_P$ = \SI{8.5e17}{\per\cubic\centi\meter} $\cdot \frac{\SI{1}{\micro\meter}}{\sigma_z}$).
However, because the charge density of the bunch was held constant in the simulations, shorter bunches caused weaker wakefields, leading to weaker beam focusing. Scalings of the simulations revealed a trade-off: reducing bunch charge by a factor of 3 necessitated quadrupling the lens' plasma density to maintain similar modulations in the energy spectrum (see Appendix \ref{appendix:A}).

Figure \ref{fig:PIC-beamlength} reveals that beam size modulations -- critical for longitudinal profile reconstruction -- persist across all electron bunches. Strikingly, these modulations remain resolvable even in the sub-femtosecond regime (red curve), enabling a direct measurement of bunch lengths deep within the attosecond regime.

\section{Conclusion}

The rapidly evolving fields of passive plasma lenses have been shown to provide the basis for a powerful diagnostic of attosecond duration electron bunches. This study demonstrates that the energy-dependent electron beam focusing in a plasma lens results in  chromatic modulation patterns that encode  the longitudinal bunch structure. An analytic model that reconstructs longitudinal bunch profiles from these modulations is presented and validated through particle-in-cell (PIC) simulations for both chirped and unchirped pulses.
Highest sensitivity depends on choosing the correct plasma lens density and length. 
The strong agreement between theory, simulation, and experiment establishes plasma lenses as  a precision diagnostic for ultrashort electron beams.

\section*{Acknowledgements}
The authors thank G. Schäfer for operating the \jeti laser system. 
This work was also funded by the Deutsche Forschungsgemeinschaft (DFG, German Research Foundation) under Project ID 392856280.

\appendix
\section{Focusing with constant gradient}
\label{appendix:C}
\begin{figure}[t]
    \centering
    \includegraphics[width=0.95\linewidth]{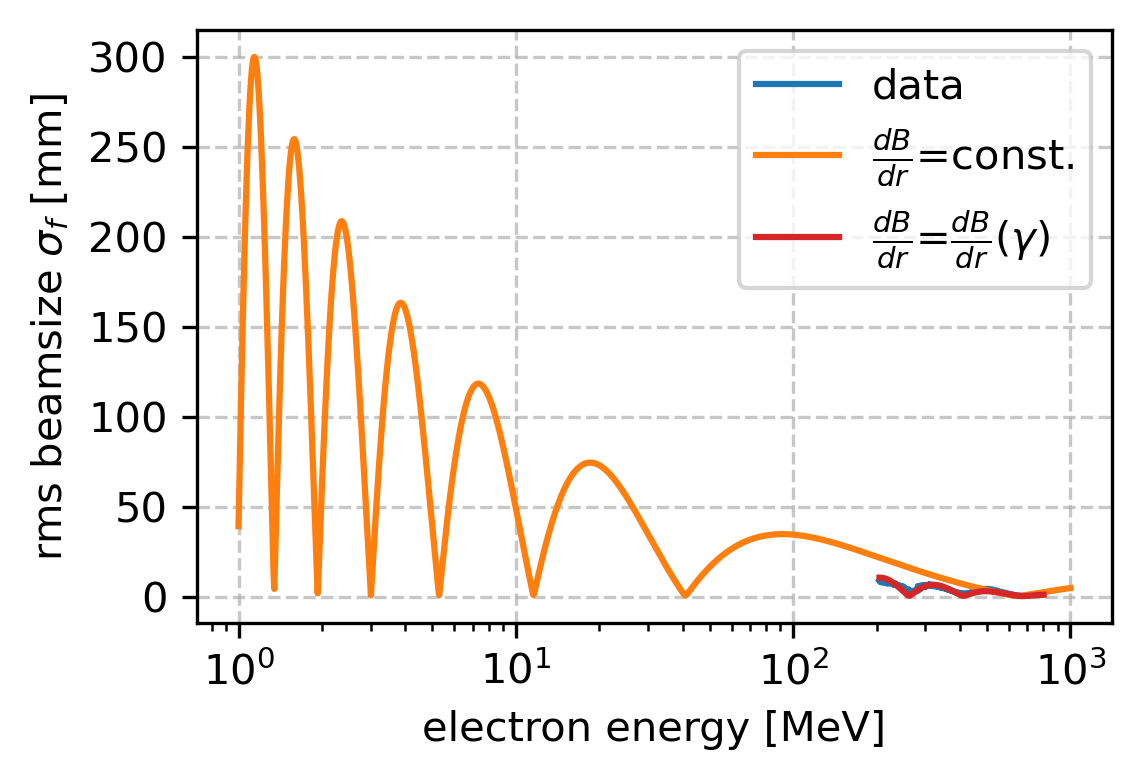}
    \caption{Root-mean-square (rms) beam size as a function of electron energy from Figure \ref{fig:experimental_results}b extended to \SI{1}{\mega\electronvolt}.}
    \label{fig:constant_gradient}
\end{figure}
Quadrupoles or solenoids can also generate modulations in the divergence and, consequently, the transverse beam size on a screen for different energies within an electron bunch (Figure \ref{fig:constant_gradient} orange curve). However, since all energies within the electron bunch experience the same gradient in this case, these modulations occur only at energies $E < \SI{50}{\mega\electronvolt}$. Figure \ref{fig:constant_gradient} extends the results of Figure \ref{fig:experimental_results}b to electron energies of $E=\SI{1}{\mega\electronvolt}$.

\section{Guidelines for Plasma Lens Optimization}
\label{appendix:A}
To facilitate experimental design, we provide scaling laws derived from simulation-validated analytic models:

\begin{itemize}
    \item \textbf{Plasma Density Limit:}
    Maximum lens density is constrained by the beam's transverse size:
    \begin{equation}
        k_p \sigma_r \approx 2 
        \label{eq:plasmalimit}
    \end{equation}

    \item \textbf{Charge-Density Trade-off:}
    Empirical scaling reveals \( Q \propto n_p^{0.5\log_2 3} \). Tripling the bunch charge (\( Q \rightarrow 3Q \)) necessitates quartering the plasma density (\( n_p \rightarrow 0.25n_p \)) to maintain equivalent focusing.

    \item \textbf{Bunch Length Compensation:}
    Doubling the bunch length (\( \sigma_z \rightarrow 2\sigma_z \)) necessitates halving the plasma wavelength (\( \lambda_p \rightarrow \lambda_p/2 \)) to maintain equivalent focusing.
\end{itemize}

These relationships coalesce into a unified scaling formula for the plasma lens length \( L \):
\begin{multline}
\label{eq:lensScaling}
    \frac{L}{\SI{2}{\milli\meter}} = \frac{\gamma/1000}{(f/\SI{40}{\milli\meter}) \cdot (Q/\SI{75}{\pico\coulomb}) \cdot (\sigma_z/\SI{1.5}{\micro\meter})^{\log_2 3}}\\
 \cdot\frac{1}{ (\sigma_r/\SI{8}{\micro\meter}) \cdot (n_p/\SI{1e17}{\per\cubic\centi\meter})^{0.5\log_2 3}}
\end{multline}

\paragraph*{Practical Example}  
For \( Q = \SI{75}{\pico\coulomb} \), \( \sigma_z = \SI{2}{\micro\meter} \), \( \gamma = 1300 \), \( n_p = \SI{4e17}{\per\cubic\centi\meter} \), and \( \sigma_r = \SI{5}{\micro\meter} \), Equation~\eqref{eq:lensScaling} predicts \( L \approx \SI{3.5}{\milli\meter} \) to achieve \( f = \SI{10}{\milli\meter} \).

\bibliography{references}

\begin{thebibliography}{28}%
\makeatletter
\providecommand \@ifxundefined [1]{%
 \@ifx{#1\undefined}
}%
\providecommand \@ifnum [1]{%
 \ifnum #1\expandafter \@firstoftwo
 \else \expandafter \@secondoftwo
 \fi
}%
\providecommand \@ifx [1]{%
 \ifx #1\expandafter \@firstoftwo
 \else \expandafter \@secondoftwo
 \fi
}%
\providecommand \natexlab [1]{#1}%
\providecommand \enquote  [1]{``#1''}%
\providecommand \bibnamefont  [1]{#1}%
\providecommand \bibfnamefont [1]{#1}%
\providecommand \citenamefont [1]{#1}%
\providecommand \href@noop [0]{\@secondoftwo}%
\providecommand \href [0]{\begingroup \@sanitize@url \@href}%
\providecommand \@href[1]{\@@startlink{#1}\@@href}%
\providecommand \@@href[1]{\endgroup#1\@@endlink}%
\providecommand \@sanitize@url [0]{\catcode `\\12\catcode `\$12\catcode
  `\&12\catcode `\#12\catcode `\^12\catcode `\_12\catcode `\%12\relax}%
\providecommand \@@startlink[1]{}%
\providecommand \@@endlink[0]{}%
\providecommand \url  [0]{\begingroup\@sanitize@url \@url }%
\providecommand \@url [1]{\endgroup\@href {#1}{\urlprefix }}%
\providecommand \urlprefix  [0]{URL }%
\providecommand \Eprint [0]{\href }%
\providecommand \doibase [0]{http://dx.doi.org/}%
\providecommand \selectlanguage [0]{\@gobble}%
\providecommand \bibinfo  [0]{\@secondoftwo}%
\providecommand \bibfield  [0]{\@secondoftwo}%
\providecommand \translation [1]{[#1]}%
\providecommand \BibitemOpen [0]{}%
\providecommand \bibitemStop [0]{}%
\providecommand \bibitemNoStop [0]{.\EOS\space}%
\providecommand \EOS [0]{\spacefactor3000\relax}%
\providecommand \BibitemShut  [1]{\csname bibitem#1\endcsname}%
\let\auto@bib@innerbib\@empty
\bibitem [{\citenamefont {Tajima}\ and\ \citenamefont
  {Dawson}(1979)}]{tajima_laser_1979}%
  \BibitemOpen
  \bibfield  {author} {\bibinfo {author} {\bibfnamefont {T.}~\bibnamefont
  {Tajima}}\ and\ \bibinfo {author} {\bibfnamefont {J.~M.}\ \bibnamefont
  {Dawson}},\ }\href {\doibase 10.1103/PhysRevLett.43.267} {\bibfield
  {journal} {\bibinfo  {journal} {Physical Review Letters}\ }\textbf {\bibinfo
  {volume} {43}},\ \bibinfo {pages} {267} (\bibinfo {year} {1979})},\ \bibinfo
  {note} {publisher: American Physical Society}\BibitemShut {NoStop}%
\bibitem [{\citenamefont {Gonsalves}\ \emph {et~al.}(2019)\citenamefont
  {Gonsalves}, \citenamefont {Nakamura}, \citenamefont {Daniels}, \citenamefont
  {Benedetti}, \citenamefont {Pieronek}, \citenamefont {de~Raadt},
  \citenamefont {Steinke}, \citenamefont {Bin}, \citenamefont {Bulanov},
  \citenamefont {van Tilborg}, \citenamefont {Geddes}, \citenamefont
  {Schroeder}, \citenamefont {Tóth}, \citenamefont {Esarey}, \citenamefont
  {Swanson}, \citenamefont {Fan-Chiang}, \citenamefont {Bagdasarov},
  \citenamefont {Bobrova}, \citenamefont {Gasilov}, \citenamefont {Korn},
  \citenamefont {Sasorov},\ and\ \citenamefont
  {Leemans}}]{gonsalves_petawatt_2019}%
  \BibitemOpen
  \bibfield  {author} {\bibinfo {author} {\bibfnamefont {A.}~\bibnamefont
  {Gonsalves}}, \bibinfo {author} {\bibfnamefont {K.}~\bibnamefont {Nakamura}},
  \bibinfo {author} {\bibfnamefont {J.}~\bibnamefont {Daniels}}, \bibinfo
  {author} {\bibfnamefont {C.}~\bibnamefont {Benedetti}}, \bibinfo {author}
  {\bibfnamefont {C.}~\bibnamefont {Pieronek}}, \bibinfo {author}
  {\bibfnamefont {T.}~\bibnamefont {de~Raadt}}, \bibinfo {author}
  {\bibfnamefont {S.}~\bibnamefont {Steinke}}, \bibinfo {author} {\bibfnamefont
  {J.}~\bibnamefont {Bin}}, \bibinfo {author} {\bibfnamefont {S.}~\bibnamefont
  {Bulanov}}, \bibinfo {author} {\bibfnamefont {J.}~\bibnamefont {van
  Tilborg}}, \bibinfo {author} {\bibfnamefont {C.}~\bibnamefont {Geddes}},
  \bibinfo {author} {\bibfnamefont {C.}~\bibnamefont {Schroeder}}, \bibinfo
  {author} {\bibfnamefont {C.}~\bibnamefont {Tóth}}, \bibinfo {author}
  {\bibfnamefont {E.}~\bibnamefont {Esarey}}, \bibinfo {author} {\bibfnamefont
  {K.}~\bibnamefont {Swanson}}, \bibinfo {author} {\bibfnamefont
  {L.}~\bibnamefont {Fan-Chiang}}, \bibinfo {author} {\bibfnamefont
  {G.}~\bibnamefont {Bagdasarov}}, \bibinfo {author} {\bibfnamefont
  {N.}~\bibnamefont {Bobrova}}, \bibinfo {author} {\bibfnamefont
  {V.}~\bibnamefont {Gasilov}}, \bibinfo {author} {\bibfnamefont
  {G.}~\bibnamefont {Korn}}, \bibinfo {author} {\bibfnamefont {P.}~\bibnamefont
  {Sasorov}}, \ and\ \bibinfo {author} {\bibfnamefont {W.}~\bibnamefont
  {Leemans}},\ }\href {\doibase 10.1103/PhysRevLett.122.084801} {\bibfield
  {journal} {\bibinfo  {journal} {Physical Review Letters}\ }\textbf {\bibinfo
  {volume} {122}},\ \bibinfo {pages} {084801} (\bibinfo {year} {2019})},\
  \bibinfo {note} {publisher: American Physical Society}\BibitemShut {NoStop}%
\bibitem [{\citenamefont {Zarini}\ \emph {et~al.}(2022)\citenamefont {Zarini},
  \citenamefont {Cabadağ}, \citenamefont {Chang}, \citenamefont {Köhler},
  \citenamefont {Kurz}, \citenamefont {Schöbel}, \citenamefont {Seidel},
  \citenamefont {Bussmann}, \citenamefont {Schramm}, \citenamefont {Irman},\
  and\ \citenamefont {Debus}}]{Zarini_BunchLength}%
  \BibitemOpen
  \bibfield  {author} {\bibinfo {author} {\bibfnamefont {O.}~\bibnamefont
  {Zarini}}, \bibinfo {author} {\bibfnamefont {J.~C.}\ \bibnamefont
  {Cabadağ}}, \bibinfo {author} {\bibfnamefont {Y.-Y.}\ \bibnamefont {Chang}},
  \bibinfo {author} {\bibfnamefont {A.}~\bibnamefont {Köhler}}, \bibinfo
  {author} {\bibfnamefont {T.}~\bibnamefont {Kurz}}, \bibinfo {author}
  {\bibfnamefont {S.}~\bibnamefont {Schöbel}}, \bibinfo {author}
  {\bibfnamefont {W.}~\bibnamefont {Seidel}}, \bibinfo {author} {\bibfnamefont
  {M.}~\bibnamefont {Bussmann}}, \bibinfo {author} {\bibfnamefont
  {U.}~\bibnamefont {Schramm}}, \bibinfo {author} {\bibfnamefont
  {A.}~\bibnamefont {Irman}}, \ and\ \bibinfo {author} {\bibfnamefont
  {A.}~\bibnamefont {Debus}},\ }\href {\doibase
  10.1103/PhysRevAccelBeams.25.012801} {\bibfield  {journal} {\bibinfo
  {journal} {Physical Review Accelerators and Beams}\ }\textbf {\bibinfo
  {volume} {25}},\ \bibinfo {pages} {12801} (\bibinfo {year}
  {2022})}\BibitemShut {NoStop}%
\bibitem [{\citenamefont {Huang}\ \emph {et~al.}(2024)\citenamefont {Huang},
  \citenamefont {Jin}, \citenamefont {Nakanii}, \citenamefont {Hosokai},\ and\
  \citenamefont {Kando}}]{huang_electro-optic_2024}%
  \BibitemOpen
  \bibfield  {author} {\bibinfo {author} {\bibfnamefont {K.}~\bibnamefont
  {Huang}}, \bibinfo {author} {\bibfnamefont {Z.}~\bibnamefont {Jin}}, \bibinfo
  {author} {\bibfnamefont {N.}~\bibnamefont {Nakanii}}, \bibinfo {author}
  {\bibfnamefont {T.}~\bibnamefont {Hosokai}}, \ and\ \bibinfo {author}
  {\bibfnamefont {M.}~\bibnamefont {Kando}},\ }\href {\doibase
  10.1038/s41377-024-01440-2} {\bibfield  {journal} {\bibinfo  {journal}
  {Light: Science \& Applications}\ }\textbf {\bibinfo {volume} {13}},\
  \bibinfo {pages} {84} (\bibinfo {year} {2024})},\ \bibinfo {note} {publisher:
  Nature Publishing Group}\BibitemShut {NoStop}%
\bibitem [{\citenamefont {Galletti}\ \emph {et~al.}(2024)\citenamefont
  {Galletti}, \citenamefont {Assmann}, \citenamefont {Couprie}, \citenamefont
  {Ferrario}, \citenamefont {Giannessi}, \citenamefont {Irman}, \citenamefont
  {Pompili},\ and\ \citenamefont {Wang}}]{galletti_prospects_2024}%
  \BibitemOpen
  \bibfield  {author} {\bibinfo {author} {\bibfnamefont {M.}~\bibnamefont
  {Galletti}}, \bibinfo {author} {\bibfnamefont {R.}~\bibnamefont {Assmann}},
  \bibinfo {author} {\bibfnamefont {M.~E.}\ \bibnamefont {Couprie}}, \bibinfo
  {author} {\bibfnamefont {M.}~\bibnamefont {Ferrario}}, \bibinfo {author}
  {\bibfnamefont {L.}~\bibnamefont {Giannessi}}, \bibinfo {author}
  {\bibfnamefont {A.}~\bibnamefont {Irman}}, \bibinfo {author} {\bibfnamefont
  {R.}~\bibnamefont {Pompili}}, \ and\ \bibinfo {author} {\bibfnamefont
  {W.}~\bibnamefont {Wang}},\ }\href {\doibase 10.1038/s41566-024-01474-3}
  {\bibfield  {journal} {\bibinfo  {journal} {Nature Photonics}\ }\textbf
  {\bibinfo {volume} {18}},\ \bibinfo {pages} {780} (\bibinfo {year} {2024})},\
  \bibinfo {note} {publisher: Nature Publishing Group}\BibitemShut {NoStop}%
\bibitem [{\citenamefont {Hidding}\ \emph {et~al.}(2023)\citenamefont
  {Hidding}, \citenamefont {Assmann}, \citenamefont {Bussmann}, \citenamefont
  {Campbell}, \citenamefont {Chang}, \citenamefont {Corde}, \citenamefont
  {Cabadağ}, \citenamefont {Debus}, \citenamefont {Döpp}, \citenamefont
  {Gilljohann}, \citenamefont {Götzfried}, \citenamefont {Foerster},
  \citenamefont {Haberstroh}, \citenamefont {Habib}, \citenamefont {Heinemann},
  \citenamefont {Hollatz}, \citenamefont {Irman}, \citenamefont {Kaluza},
  \citenamefont {Karsch}, \citenamefont {Kononenko}, \citenamefont {Knetsch},
  \citenamefont {Kurz}, \citenamefont {Kuschel}, \citenamefont {Köhler},
  \citenamefont {Ossa}, \citenamefont {Nutter}, \citenamefont {Pausch},
  \citenamefont {Raj}, \citenamefont {Schramm}, \citenamefont {Schöbel},
  \citenamefont {Seidel}, \citenamefont {Steiniger}, \citenamefont {Ufer},
  \citenamefont {Yeung}, \citenamefont {Zarini},\ and\ \citenamefont
  {Zepf}}]{hidding2023}%
  \BibitemOpen
  \bibfield  {author} {\bibinfo {author} {\bibfnamefont {B.}~\bibnamefont
  {Hidding}}, \bibinfo {author} {\bibfnamefont {R.}~\bibnamefont {Assmann}},
  \bibinfo {author} {\bibfnamefont {M.}~\bibnamefont {Bussmann}}, \bibinfo
  {author} {\bibfnamefont {D.}~\bibnamefont {Campbell}}, \bibinfo {author}
  {\bibfnamefont {Y.-Y.}\ \bibnamefont {Chang}}, \bibinfo {author}
  {\bibfnamefont {S.}~\bibnamefont {Corde}}, \bibinfo {author} {\bibfnamefont
  {J.~C.}\ \bibnamefont {Cabadağ}}, \bibinfo {author} {\bibfnamefont
  {A.}~\bibnamefont {Debus}}, \bibinfo {author} {\bibfnamefont
  {A.}~\bibnamefont {Döpp}}, \bibinfo {author} {\bibfnamefont
  {M.}~\bibnamefont {Gilljohann}}, \bibinfo {author} {\bibfnamefont
  {J.}~\bibnamefont {Götzfried}}, \bibinfo {author} {\bibfnamefont {F.~M.}\
  \bibnamefont {Foerster}}, \bibinfo {author} {\bibfnamefont {F.}~\bibnamefont
  {Haberstroh}}, \bibinfo {author} {\bibfnamefont {F.}~\bibnamefont {Habib}},
  \bibinfo {author} {\bibfnamefont {T.}~\bibnamefont {Heinemann}}, \bibinfo
  {author} {\bibfnamefont {D.}~\bibnamefont {Hollatz}}, \bibinfo {author}
  {\bibfnamefont {A.}~\bibnamefont {Irman}}, \bibinfo {author} {\bibfnamefont
  {M.}~\bibnamefont {Kaluza}}, \bibinfo {author} {\bibfnamefont
  {S.}~\bibnamefont {Karsch}}, \bibinfo {author} {\bibfnamefont
  {O.}~\bibnamefont {Kononenko}}, \bibinfo {author} {\bibfnamefont
  {A.}~\bibnamefont {Knetsch}}, \bibinfo {author} {\bibfnamefont
  {T.}~\bibnamefont {Kurz}}, \bibinfo {author} {\bibfnamefont {S.}~\bibnamefont
  {Kuschel}}, \bibinfo {author} {\bibfnamefont {A.}~\bibnamefont {Köhler}},
  \bibinfo {author} {\bibfnamefont {A.~M. d.~l.}\ \bibnamefont {Ossa}},
  \bibinfo {author} {\bibfnamefont {A.}~\bibnamefont {Nutter}}, \bibinfo
  {author} {\bibfnamefont {R.}~\bibnamefont {Pausch}}, \bibinfo {author}
  {\bibfnamefont {G.}~\bibnamefont {Raj}}, \bibinfo {author} {\bibfnamefont
  {U.}~\bibnamefont {Schramm}}, \bibinfo {author} {\bibfnamefont
  {S.}~\bibnamefont {Schöbel}}, \bibinfo {author} {\bibfnamefont
  {A.}~\bibnamefont {Seidel}}, \bibinfo {author} {\bibfnamefont
  {K.}~\bibnamefont {Steiniger}}, \bibinfo {author} {\bibfnamefont
  {P.}~\bibnamefont {Ufer}}, \bibinfo {author} {\bibfnamefont {M.}~\bibnamefont
  {Yeung}}, \bibinfo {author} {\bibfnamefont {O.}~\bibnamefont {Zarini}}, \
  and\ \bibinfo {author} {\bibfnamefont {M.}~\bibnamefont {Zepf}},\ }\href
  {\doibase 10.3390/photonics10020099} {\bibfield  {journal} {\bibinfo
  {journal} {Photonics}\ }\textbf {\bibinfo {volume} {10}} (\bibinfo {year}
  {2023}),\ 10.3390/photonics10020099}\BibitemShut {NoStop}%
\bibitem [{\citenamefont {Foster}\ \emph {et~al.}(2023)\citenamefont {Foster},
  \citenamefont {D’Arcy},\ and\ \citenamefont
  {Lindstrøm}}]{foster_hybrid_2023}%
  \BibitemOpen
  \bibfield  {author} {\bibinfo {author} {\bibfnamefont {B.}~\bibnamefont
  {Foster}}, \bibinfo {author} {\bibfnamefont {R.}~\bibnamefont {D’Arcy}}, \
  and\ \bibinfo {author} {\bibfnamefont {C.~A.}\ \bibnamefont {Lindstrøm}},\
  }\href {\doibase 10.1088/1367-2630/acf395} {\bibfield  {journal} {\bibinfo
  {journal} {New Journal of Physics}\ }\textbf {\bibinfo {volume} {25}},\
  \bibinfo {pages} {093037} (\bibinfo {year} {2023})},\ \bibinfo {note}
  {publisher: IOP Publishing}\BibitemShut {NoStop}%
\bibitem [{\citenamefont {Antipov}\ \emph {et~al.}(2021)\citenamefont
  {Antipov}, \citenamefont {Pousa}, \citenamefont {Agapov}, \citenamefont
  {Brinkmann}, \citenamefont {Maier}, \citenamefont {Jalas}, \citenamefont
  {Jeppe}, \citenamefont {Kirchen}, \citenamefont {Leemans}, \citenamefont
  {de~la Ossa}, \citenamefont {Osterhoff}, \citenamefont {Th\'evenet},\ and\
  \citenamefont {Winkler}}]{antipov2021}%
  \BibitemOpen
  \bibfield  {author} {\bibinfo {author} {\bibfnamefont {S.~A.}\ \bibnamefont
  {Antipov}}, \bibinfo {author} {\bibfnamefont {A.~F.}\ \bibnamefont {Pousa}},
  \bibinfo {author} {\bibfnamefont {I.}~\bibnamefont {Agapov}}, \bibinfo
  {author} {\bibfnamefont {R.}~\bibnamefont {Brinkmann}}, \bibinfo {author}
  {\bibfnamefont {A.~R.}\ \bibnamefont {Maier}}, \bibinfo {author}
  {\bibfnamefont {S.}~\bibnamefont {Jalas}}, \bibinfo {author} {\bibfnamefont
  {L.}~\bibnamefont {Jeppe}}, \bibinfo {author} {\bibfnamefont
  {M.}~\bibnamefont {Kirchen}}, \bibinfo {author} {\bibfnamefont {W.~P.}\
  \bibnamefont {Leemans}}, \bibinfo {author} {\bibfnamefont {A.~M.}\
  \bibnamefont {de~la Ossa}}, \bibinfo {author} {\bibfnamefont
  {J.}~\bibnamefont {Osterhoff}}, \bibinfo {author} {\bibfnamefont
  {M.}~\bibnamefont {Th\'evenet}}, \ and\ \bibinfo {author} {\bibfnamefont
  {P.}~\bibnamefont {Winkler}},\ }\href {\doibase
  10.1103/PhysRevAccelBeams.24.111301} {\bibfield  {journal} {\bibinfo
  {journal} {Phys. Rev. Accel. Beams}\ }\textbf {\bibinfo {volume} {24}},\
  \bibinfo {pages} {111301} (\bibinfo {year} {2021})}\BibitemShut {NoStop}%
\bibitem [{\citenamefont {Behrens}\ \emph {et~al.}(2014)\citenamefont
  {Behrens}, \citenamefont {Decker}, \citenamefont {Ding}, \citenamefont
  {Dolgashev}, \citenamefont {Frisch}, \citenamefont {Huang}, \citenamefont
  {Krejcik}, \citenamefont {Loos}, \citenamefont {Lutman}, \citenamefont
  {Maxwell}, \citenamefont {Turner}, \citenamefont {Wang}, \citenamefont
  {Wang}, \citenamefont {Welch},\ and\ \citenamefont {Wu}}]{Behrens2014}%
  \BibitemOpen
  \bibfield  {author} {\bibinfo {author} {\bibfnamefont {C.}~\bibnamefont
  {Behrens}}, \bibinfo {author} {\bibfnamefont {F.-J.}\ \bibnamefont {Decker}},
  \bibinfo {author} {\bibfnamefont {Y.}~\bibnamefont {Ding}}, \bibinfo {author}
  {\bibfnamefont {V.~A.}\ \bibnamefont {Dolgashev}}, \bibinfo {author}
  {\bibfnamefont {J.}~\bibnamefont {Frisch}}, \bibinfo {author} {\bibfnamefont
  {Z.}~\bibnamefont {Huang}}, \bibinfo {author} {\bibfnamefont
  {P.}~\bibnamefont {Krejcik}}, \bibinfo {author} {\bibfnamefont
  {H.}~\bibnamefont {Loos}}, \bibinfo {author} {\bibfnamefont {A.}~\bibnamefont
  {Lutman}}, \bibinfo {author} {\bibfnamefont {T.~J.}\ \bibnamefont {Maxwell}},
  \bibinfo {author} {\bibfnamefont {J.}~\bibnamefont {Turner}}, \bibinfo
  {author} {\bibfnamefont {J.}~\bibnamefont {Wang}}, \bibinfo {author}
  {\bibfnamefont {M.-H.}\ \bibnamefont {Wang}}, \bibinfo {author}
  {\bibfnamefont {J.}~\bibnamefont {Welch}}, \ and\ \bibinfo {author}
  {\bibfnamefont {J.}~\bibnamefont {Wu}},\ }\href {\doibase 10.1038/ncomms4762}
  {\bibfield  {journal} {\bibinfo  {journal} {Nature Communications}\ }\textbf
  {\bibinfo {volume} {5}},\ \bibinfo {pages} {3762} (\bibinfo {year}
  {2014})}\BibitemShut {NoStop}%
\bibitem [{\citenamefont {Dolgashev}\ \emph {et~al.}(2014)\citenamefont
  {Dolgashev}, \citenamefont {Bowden}, \citenamefont {Ding}, \citenamefont
  {Emma}, \citenamefont {Krejcik}, \citenamefont {Lewandowski}, \citenamefont
  {Limborg}, \citenamefont {Litos}, \citenamefont {Wang},\ and\ \citenamefont
  {Xiang}}]{Dolgashev_2014_x-band}%
  \BibitemOpen
  \bibfield  {author} {\bibinfo {author} {\bibfnamefont {V.~A.}\ \bibnamefont
  {Dolgashev}}, \bibinfo {author} {\bibfnamefont {G.}~\bibnamefont {Bowden}},
  \bibinfo {author} {\bibfnamefont {Y.}~\bibnamefont {Ding}}, \bibinfo {author}
  {\bibfnamefont {P.}~\bibnamefont {Emma}}, \bibinfo {author} {\bibfnamefont
  {P.}~\bibnamefont {Krejcik}}, \bibinfo {author} {\bibfnamefont
  {J.}~\bibnamefont {Lewandowski}}, \bibinfo {author} {\bibfnamefont
  {C.}~\bibnamefont {Limborg}}, \bibinfo {author} {\bibfnamefont
  {M.}~\bibnamefont {Litos}}, \bibinfo {author} {\bibfnamefont
  {J.}~\bibnamefont {Wang}}, \ and\ \bibinfo {author} {\bibfnamefont
  {D.}~\bibnamefont {Xiang}},\ }\href {\doibase 10.1103/PhysRevSTAB.17.102801}
  {\bibfield  {journal} {\bibinfo  {journal} {Physical Review Special Topics -
  Accelerators and Beams}\ }\textbf {\bibinfo {volume} {17}},\ \bibinfo {pages}
  {102801} (\bibinfo {year} {2014})}\BibitemShut {NoStop}%
\bibitem [{\citenamefont {Maxson}\ \emph {et~al.}(2017)\citenamefont {Maxson},
  \citenamefont {Cesar}, \citenamefont {Calmasini}, \citenamefont {Ody},
  \citenamefont {Musumeci},\ and\ \citenamefont
  {Alesini}}]{maxson_direct_2017}%
  \BibitemOpen
  \bibfield  {author} {\bibinfo {author} {\bibfnamefont {J.}~\bibnamefont
  {Maxson}}, \bibinfo {author} {\bibfnamefont {D.}~\bibnamefont {Cesar}},
  \bibinfo {author} {\bibfnamefont {G.}~\bibnamefont {Calmasini}}, \bibinfo
  {author} {\bibfnamefont {A.}~\bibnamefont {Ody}}, \bibinfo {author}
  {\bibfnamefont {P.}~\bibnamefont {Musumeci}}, \ and\ \bibinfo {author}
  {\bibfnamefont {D.}~\bibnamefont {Alesini}},\ }\href {\doibase
  10.1103/PhysRevLett.118.154802} {\bibfield  {journal} {\bibinfo  {journal}
  {Physical Review Letters}\ }\textbf {\bibinfo {volume} {118}},\ \bibinfo
  {pages} {154802} (\bibinfo {year} {2017})},\ \bibinfo {note} {publisher:
  American Physical Society}\BibitemShut {NoStop}%
\bibitem [{\citenamefont {Buck}\ \emph {et~al.}(2011)\citenamefont {Buck},
  \citenamefont {Nicolai}, \citenamefont {Schmid}, \citenamefont {Sears},
  \citenamefont {Sävert}, \citenamefont {Mikhailova}, \citenamefont {Krausz},
  \citenamefont {Kaluza},\ and\ \citenamefont
  {Veisz}}]{Buck_2011_ProbingFaradayElectronbunch}%
  \BibitemOpen
  \bibfield  {author} {\bibinfo {author} {\bibfnamefont {A.}~\bibnamefont
  {Buck}}, \bibinfo {author} {\bibfnamefont {M.}~\bibnamefont {Nicolai}},
  \bibinfo {author} {\bibfnamefont {K.}~\bibnamefont {Schmid}}, \bibinfo
  {author} {\bibfnamefont {C.~M.~S.}\ \bibnamefont {Sears}}, \bibinfo {author}
  {\bibfnamefont {A.}~\bibnamefont {Sävert}}, \bibinfo {author} {\bibfnamefont
  {J.~M.}\ \bibnamefont {Mikhailova}}, \bibinfo {author} {\bibfnamefont
  {F.}~\bibnamefont {Krausz}}, \bibinfo {author} {\bibfnamefont {M.~C.}\
  \bibnamefont {Kaluza}}, \ and\ \bibinfo {author} {\bibfnamefont
  {L.}~\bibnamefont {Veisz}},\ }\href {\doibase 10.1038/nphys1942} {\bibfield
  {journal} {\bibinfo  {journal} {Nature Physics}\ }\textbf {\bibinfo {volume}
  {7}},\ \bibinfo {pages} {543} (\bibinfo {year} {2011})}\BibitemShut {NoStop}%
\bibitem [{\citenamefont {Barber}\ \emph {et~al.}(2017)\citenamefont {Barber},
  \citenamefont {Tilborg}, \citenamefont {Schroeder}, \citenamefont {Lehe},
  \citenamefont {Tsai}, \citenamefont {Swanson}, \citenamefont {Steinke},
  \citenamefont {Nakamura}, \citenamefont {Geddes}, \citenamefont {Benedetti},
  \citenamefont {Esarey},\ and\ \citenamefont {Leemans}}]{Barber2017}%
  \BibitemOpen
  \bibfield  {author} {\bibinfo {author} {\bibfnamefont {S.~K.}\ \bibnamefont
  {Barber}}, \bibinfo {author} {\bibfnamefont {J.~V.}\ \bibnamefont {Tilborg}},
  \bibinfo {author} {\bibfnamefont {C.~B.}\ \bibnamefont {Schroeder}}, \bibinfo
  {author} {\bibfnamefont {R.}~\bibnamefont {Lehe}}, \bibinfo {author}
  {\bibfnamefont {H.~E.}\ \bibnamefont {Tsai}}, \bibinfo {author}
  {\bibfnamefont {K.~K.}\ \bibnamefont {Swanson}}, \bibinfo {author}
  {\bibfnamefont {S.}~\bibnamefont {Steinke}}, \bibinfo {author} {\bibfnamefont
  {K.}~\bibnamefont {Nakamura}}, \bibinfo {author} {\bibfnamefont {C.~G.}\
  \bibnamefont {Geddes}}, \bibinfo {author} {\bibfnamefont {C.}~\bibnamefont
  {Benedetti}}, \bibinfo {author} {\bibfnamefont {E.}~\bibnamefont {Esarey}}, \
  and\ \bibinfo {author} {\bibfnamefont {W.~P.}\ \bibnamefont {Leemans}},\
  }\href {\doibase 10.1103/PhysRevLett.119.104801} {\bibfield  {journal}
  {\bibinfo  {journal} {Physical Review Letters}\ }\textbf {\bibinfo {volume}
  {119}},\ \bibinfo {pages} {1} (\bibinfo {year} {2017})}\BibitemShut {NoStop}%
\bibitem [{\citenamefont {Lindstrøm}\ \emph {et~al.}(2018)\citenamefont
  {Lindstrøm}, \citenamefont {Adli}, \citenamefont {Boyle}, \citenamefont
  {Corsini}, \citenamefont {Dyson}, \citenamefont {Farabolini}, \citenamefont
  {Hooker}, \citenamefont {Meisel}, \citenamefont {Osterhoff}, \citenamefont
  {Röckemann}, \citenamefont {Schaper},\ and\ \citenamefont
  {Sjobak}}]{Lindstroem_2018_emittance}%
  \BibitemOpen
  \bibfield  {author} {\bibinfo {author} {\bibfnamefont {C.}~\bibnamefont
  {Lindstrøm}}, \bibinfo {author} {\bibfnamefont {E.}~\bibnamefont {Adli}},
  \bibinfo {author} {\bibfnamefont {G.}~\bibnamefont {Boyle}}, \bibinfo
  {author} {\bibfnamefont {R.}~\bibnamefont {Corsini}}, \bibinfo {author}
  {\bibfnamefont {A.}~\bibnamefont {Dyson}}, \bibinfo {author} {\bibfnamefont
  {W.}~\bibnamefont {Farabolini}}, \bibinfo {author} {\bibfnamefont
  {S.}~\bibnamefont {Hooker}}, \bibinfo {author} {\bibfnamefont
  {M.}~\bibnamefont {Meisel}}, \bibinfo {author} {\bibfnamefont
  {J.}~\bibnamefont {Osterhoff}}, \bibinfo {author} {\bibfnamefont {J.-H.}\
  \bibnamefont {Röckemann}}, \bibinfo {author} {\bibfnamefont
  {L.}~\bibnamefont {Schaper}}, \ and\ \bibinfo {author} {\bibfnamefont
  {K.}~\bibnamefont {Sjobak}},\ }\href {\doibase
  10.1103/PhysRevLett.121.194801} {\bibfield  {journal} {\bibinfo  {journal}
  {Physical Review Letters}\ }\textbf {\bibinfo {volume} {121}},\ \bibinfo
  {pages} {194801} (\bibinfo {year} {2018})}\BibitemShut {NoStop}%
\bibitem [{\citenamefont {Kuschel}\ \emph {et~al.}(2016)\citenamefont
  {Kuschel}, \citenamefont {Hollatz}, \citenamefont {Heinemann}, \citenamefont
  {Karger}, \citenamefont {Schwab}, \citenamefont {Ullmann}, \citenamefont
  {Knetsch}, \citenamefont {Seidel}, \citenamefont {Rödel}, \citenamefont
  {Yeung}, \citenamefont {Leier}, \citenamefont {Blinne}, \citenamefont {Ding},
  \citenamefont {Kurz}, \citenamefont {Corvan}, \citenamefont {Sävert},
  \citenamefont {Karsch}, \citenamefont {Kaluza}, \citenamefont {Hidding},\
  and\ \citenamefont {Zepf}}]{Kuschel_passiveLens}%
  \BibitemOpen
  \bibfield  {author} {\bibinfo {author} {\bibfnamefont {S.}~\bibnamefont
  {Kuschel}}, \bibinfo {author} {\bibfnamefont {D.}~\bibnamefont {Hollatz}},
  \bibinfo {author} {\bibfnamefont {T.}~\bibnamefont {Heinemann}}, \bibinfo
  {author} {\bibfnamefont {O.}~\bibnamefont {Karger}}, \bibinfo {author}
  {\bibfnamefont {M.}~\bibnamefont {Schwab}}, \bibinfo {author} {\bibfnamefont
  {D.}~\bibnamefont {Ullmann}}, \bibinfo {author} {\bibfnamefont
  {A.}~\bibnamefont {Knetsch}}, \bibinfo {author} {\bibfnamefont
  {A.}~\bibnamefont {Seidel}}, \bibinfo {author} {\bibfnamefont
  {C.}~\bibnamefont {Rödel}}, \bibinfo {author} {\bibfnamefont
  {M.}~\bibnamefont {Yeung}}, \bibinfo {author} {\bibfnamefont
  {M.}~\bibnamefont {Leier}}, \bibinfo {author} {\bibfnamefont
  {A.}~\bibnamefont {Blinne}}, \bibinfo {author} {\bibfnamefont
  {H.}~\bibnamefont {Ding}}, \bibinfo {author} {\bibfnamefont {T.}~\bibnamefont
  {Kurz}}, \bibinfo {author} {\bibfnamefont {D.}~\bibnamefont {Corvan}},
  \bibinfo {author} {\bibfnamefont {A.}~\bibnamefont {Sävert}}, \bibinfo
  {author} {\bibfnamefont {S.}~\bibnamefont {Karsch}}, \bibinfo {author}
  {\bibfnamefont {M.}~\bibnamefont {Kaluza}}, \bibinfo {author} {\bibfnamefont
  {B.}~\bibnamefont {Hidding}}, \ and\ \bibinfo {author} {\bibfnamefont
  {M.}~\bibnamefont {Zepf}},\ }\href {\doibase
  10.1103/PhysRevAccelBeams.19.071301} {\bibfield  {journal} {\bibinfo
  {journal} {Physical Review Accelerators and Beams}\ }\textbf {\bibinfo
  {volume} {19}},\ \bibinfo {pages} {71301} (\bibinfo {year}
  {2016})}\BibitemShut {NoStop}%
\bibitem [{\citenamefont {Thaury}\ \emph {et~al.}(2015)\citenamefont {Thaury},
  \citenamefont {Guillaume}, \citenamefont {Döpp}, \citenamefont {Lehe},
  \citenamefont {Lifschitz}, \citenamefont {Ta~Phuoc}, \citenamefont {Gautier},
  \citenamefont {Goddet}, \citenamefont {Tafzi}, \citenamefont {Flacco},
  \citenamefont {Tissandier}, \citenamefont {Sebban}, \citenamefont {Rousse},\
  and\ \citenamefont {Malka}}]{Thaury2015}%
  \BibitemOpen
  \bibfield  {author} {\bibinfo {author} {\bibfnamefont {C.}~\bibnamefont
  {Thaury}}, \bibinfo {author} {\bibfnamefont {E.}~\bibnamefont {Guillaume}},
  \bibinfo {author} {\bibfnamefont {A.}~\bibnamefont {Döpp}}, \bibinfo
  {author} {\bibfnamefont {R.}~\bibnamefont {Lehe}}, \bibinfo {author}
  {\bibfnamefont {A.}~\bibnamefont {Lifschitz}}, \bibinfo {author}
  {\bibfnamefont {K.}~\bibnamefont {Ta~Phuoc}}, \bibinfo {author}
  {\bibfnamefont {J.}~\bibnamefont {Gautier}}, \bibinfo {author} {\bibfnamefont
  {J.~P.}\ \bibnamefont {Goddet}}, \bibinfo {author} {\bibfnamefont
  {A.}~\bibnamefont {Tafzi}}, \bibinfo {author} {\bibfnamefont
  {A.}~\bibnamefont {Flacco}}, \bibinfo {author} {\bibfnamefont
  {F.}~\bibnamefont {Tissandier}}, \bibinfo {author} {\bibfnamefont
  {S.}~\bibnamefont {Sebban}}, \bibinfo {author} {\bibfnamefont
  {A.}~\bibnamefont {Rousse}}, \ and\ \bibinfo {author} {\bibfnamefont
  {V.}~\bibnamefont {Malka}},\ }\href {\doibase 10.1038/ncomms7860} {\bibfield
  {journal} {\bibinfo  {journal} {Nature Communications}\ }\textbf {\bibinfo
  {volume} {6}} (\bibinfo {year} {2015}),\ 10.1038/ncomms7860}\BibitemShut
  {NoStop}%
\bibitem [{\citenamefont {Doss}\ \emph {et~al.}(2019)\citenamefont {Doss},
  \citenamefont {Adli}, \citenamefont {Ariniello}, \citenamefont {Cary},
  \citenamefont {Corde}, \citenamefont {Hidding}, \citenamefont {Hogan},
  \citenamefont {Hunt-Stone}, \citenamefont {Joshi}, \citenamefont {Marsh},
  \citenamefont {Rosenzweig}, \citenamefont {Vafaei-Najafabadi}, \citenamefont
  {Yakimenko},\ and\ \citenamefont {Litos}}]{doss_laser-ionized_2019}%
  \BibitemOpen
  \bibfield  {author} {\bibinfo {author} {\bibfnamefont {C.}~\bibnamefont
  {Doss}}, \bibinfo {author} {\bibfnamefont {E.}~\bibnamefont {Adli}}, \bibinfo
  {author} {\bibfnamefont {R.}~\bibnamefont {Ariniello}}, \bibinfo {author}
  {\bibfnamefont {J.}~\bibnamefont {Cary}}, \bibinfo {author} {\bibfnamefont
  {S.}~\bibnamefont {Corde}}, \bibinfo {author} {\bibfnamefont
  {B.}~\bibnamefont {Hidding}}, \bibinfo {author} {\bibfnamefont
  {M.}~\bibnamefont {Hogan}}, \bibinfo {author} {\bibfnamefont
  {K.}~\bibnamefont {Hunt-Stone}}, \bibinfo {author} {\bibfnamefont
  {C.}~\bibnamefont {Joshi}}, \bibinfo {author} {\bibfnamefont
  {K.}~\bibnamefont {Marsh}}, \bibinfo {author} {\bibfnamefont
  {J.}~\bibnamefont {Rosenzweig}}, \bibinfo {author} {\bibfnamefont
  {N.}~\bibnamefont {Vafaei-Najafabadi}}, \bibinfo {author} {\bibfnamefont
  {V.}~\bibnamefont {Yakimenko}}, \ and\ \bibinfo {author} {\bibfnamefont
  {M.}~\bibnamefont {Litos}},\ }\href {\doibase
  10.1103/PhysRevAccelBeams.22.111001} {\bibfield  {journal} {\bibinfo
  {journal} {Physical Review Accelerators and Beams}\ }\textbf {\bibinfo
  {volume} {22}},\ \bibinfo {pages} {111001} (\bibinfo {year} {2019})},\
  \bibinfo {note} {publisher: American Physical Society}\BibitemShut {NoStop}%
\bibitem [{\citenamefont {Doss}\ \emph {et~al.}(2023)\citenamefont {Doss},
  \citenamefont {Ariniello}, \citenamefont {Cary}, \citenamefont {Corde},
  \citenamefont {Ekerfelt}, \citenamefont {Gerstmayr}, \citenamefont {Gessner},
  \citenamefont {Gilljohann}, \citenamefont {Hansel}, \citenamefont {Hidding},
  \citenamefont {Hogan}, \citenamefont {Knetsch}, \citenamefont {Lee},
  \citenamefont {Marsh}, \citenamefont {O’Shea}, \citenamefont {San
  Miguel~Claveria}, \citenamefont {Storey}, \citenamefont {Sutherland},
  \citenamefont {Zhang},\ and\ \citenamefont {Litos}}]{doss_underdense_2023}%
  \BibitemOpen
  \bibfield  {author} {\bibinfo {author} {\bibfnamefont {C.}~\bibnamefont
  {Doss}}, \bibinfo {author} {\bibfnamefont {R.}~\bibnamefont {Ariniello}},
  \bibinfo {author} {\bibfnamefont {J.}~\bibnamefont {Cary}}, \bibinfo {author}
  {\bibfnamefont {S.}~\bibnamefont {Corde}}, \bibinfo {author} {\bibfnamefont
  {H.}~\bibnamefont {Ekerfelt}}, \bibinfo {author} {\bibfnamefont
  {E.}~\bibnamefont {Gerstmayr}}, \bibinfo {author} {\bibfnamefont
  {S.}~\bibnamefont {Gessner}}, \bibinfo {author} {\bibfnamefont
  {M.}~\bibnamefont {Gilljohann}}, \bibinfo {author} {\bibfnamefont
  {C.}~\bibnamefont {Hansel}}, \bibinfo {author} {\bibfnamefont
  {B.}~\bibnamefont {Hidding}}, \bibinfo {author} {\bibfnamefont
  {M.}~\bibnamefont {Hogan}}, \bibinfo {author} {\bibfnamefont
  {A.}~\bibnamefont {Knetsch}}, \bibinfo {author} {\bibfnamefont
  {V.}~\bibnamefont {Lee}}, \bibinfo {author} {\bibfnamefont {K.}~\bibnamefont
  {Marsh}}, \bibinfo {author} {\bibfnamefont {B.}~\bibnamefont {O’Shea}},
  \bibinfo {author} {\bibfnamefont {P.}~\bibnamefont {San Miguel~Claveria}},
  \bibinfo {author} {\bibfnamefont {D.}~\bibnamefont {Storey}}, \bibinfo
  {author} {\bibfnamefont {A.}~\bibnamefont {Sutherland}}, \bibinfo {author}
  {\bibfnamefont {C.}~\bibnamefont {Zhang}}, \ and\ \bibinfo {author}
  {\bibfnamefont {M.}~\bibnamefont {Litos}},\ }\href {\doibase
  10.1103/PhysRevAccelBeams.26.031302} {\bibfield  {journal} {\bibinfo
  {journal} {Physical Review Accelerators and Beams}\ }\textbf {\bibinfo
  {volume} {26}},\ \bibinfo {pages} {031302} (\bibinfo {year} {2023})},\
  \bibinfo {note} {publisher: American Physical Society}\BibitemShut {NoStop}%
\bibitem [{\citenamefont {Keinigs}\ and\ \citenamefont
  {Jones}(1987)}]{Keinigs_1987_PlasmawaveTheory}%
  \BibitemOpen
  \bibfield  {author} {\bibinfo {author} {\bibfnamefont {R.}~\bibnamefont
  {Keinigs}}\ and\ \bibinfo {author} {\bibfnamefont {M.~E.}\ \bibnamefont
  {Jones}},\ }\href {\doibase 10.1063/1.866183} {\bibfield  {journal} {\bibinfo
   {journal} {The Physics of Fluids}\ }\textbf {\bibinfo {volume} {30}},\
  \bibinfo {pages} {252} (\bibinfo {year} {1987})}\BibitemShut {NoStop}%
\bibitem [{\citenamefont {Verra}(2022)}]{Vera_2022_wakefield-theory}%
  \BibitemOpen
  \bibfield  {author} {\bibinfo {author} {\bibfnamefont {L.}~\bibnamefont
  {Verra}},\ }\href@noop {} {\  (\bibinfo {year} {2022})}\BibitemShut {NoStop}%
\bibitem [{\citenamefont {Allen}\ \emph {et~al.}(2012)\citenamefont {Allen},
  \citenamefont {Yakimenko}, \citenamefont {Babzien}, \citenamefont {Fedurin},
  \citenamefont {Kusche},\ and\ \citenamefont
  {Muggli}}]{Allen_2012_fillamentation}%
  \BibitemOpen
  \bibfield  {author} {\bibinfo {author} {\bibfnamefont {B.}~\bibnamefont
  {Allen}}, \bibinfo {author} {\bibfnamefont {V.}~\bibnamefont {Yakimenko}},
  \bibinfo {author} {\bibfnamefont {M.}~\bibnamefont {Babzien}}, \bibinfo
  {author} {\bibfnamefont {M.}~\bibnamefont {Fedurin}}, \bibinfo {author}
  {\bibfnamefont {K.}~\bibnamefont {Kusche}}, \ and\ \bibinfo {author}
  {\bibfnamefont {P.}~\bibnamefont {Muggli}},\ }\href {\doibase
  10.1103/PhysRevLett.109.185007} {\bibfield  {journal} {\bibinfo  {journal}
  {Physical Review Letters}\ }\textbf {\bibinfo {volume} {109}},\ \bibinfo
  {pages} {185007} (\bibinfo {year} {2012})}\BibitemShut {NoStop}%
\bibitem [{\citenamefont {Esarey}\ and\ \citenamefont
  {Pilloff}(1995)}]{Esarey_1995_separatrix}%
  \BibitemOpen
  \bibfield  {author} {\bibinfo {author} {\bibfnamefont {E.}~\bibnamefont
  {Esarey}}\ and\ \bibinfo {author} {\bibfnamefont {M.}~\bibnamefont
  {Pilloff}},\ }\href {\doibase 10.1063/1.871358} {\bibfield  {journal}
  {\bibinfo  {journal} {Physics of Plasmas}\ }\textbf {\bibinfo {volume} {2}},\
  \bibinfo {pages} {1432} (\bibinfo {year} {1995})}\BibitemShut {NoStop}%
\bibitem [{\citenamefont {Reiser}(2008)}]{Reiser_2008_ElectronBasic}%
  \BibitemOpen
  \bibfield  {author} {\bibinfo {author} {\bibfnamefont {M.}~\bibnamefont
  {Reiser}},\ }\href {\doibase 10.1002/9783527622047} {\emph {\bibinfo {title}
  {Theory and Design of Charged Particle Beams}}}\ (\bibinfo  {publisher}
  {Wiley},\ \bibinfo {year} {2008})\BibitemShut {NoStop}%
\bibitem [{\citenamefont {Courant}\ and\ \citenamefont
  {Snyder}(1958)}]{Courant_1958_courant-snyder-paramter}%
  \BibitemOpen
  \bibfield  {author} {\bibinfo {author} {\bibfnamefont {E.~D.}\ \bibnamefont
  {Courant}}\ and\ \bibinfo {author} {\bibfnamefont {H.~S.}\ \bibnamefont
  {Snyder}},\ }\href {\doibase https://doi.org/10.1016/0003-4916(58)90012-5}
  {\bibfield  {journal} {\bibinfo  {journal} {Annals of Physics}\ }\textbf
  {\bibinfo {volume} {3}},\ \bibinfo {pages} {1} (\bibinfo {year}
  {1958})}\BibitemShut {NoStop}%
\bibitem [{\citenamefont {{ANSYS, Inc.}}(2019)}]{ANSYSFluentGuide2019}%
  \BibitemOpen
  \bibfield  {author} {\bibinfo {author} {\bibnamefont {{ANSYS, Inc.}}},\
  }\href@noop {} {\emph {\bibinfo {title} {ANSYS Fluent User's Guide}}}
  (\bibinfo {year} {2019}),\ \bibinfo {note} {accessed on [Date]}\BibitemShut
  {NoStop}%
\bibitem [{\citenamefont {Hüther}(2015)}]{Hüther_gasjet_density}%
  \BibitemOpen
  \bibfield  {author} {\bibinfo {author} {\bibfnamefont {M.~J.}\ \bibnamefont
  {Hüther}},\ }\href@noop {} {\enquote {\bibinfo {title} {Design and
  characterisation of supersonic nozzles for shock front electron injection in
  laser wakefield acceleration},}\ } (\bibinfo {year} {2015})\BibitemShut
  {NoStop}%
\bibitem [{\citenamefont {Kurz}\ \emph {et~al.}(2018)\citenamefont {Kurz},
  \citenamefont {Couperus}, \citenamefont {Krämer}, \citenamefont {Ding},
  \citenamefont {Kuschel}, \citenamefont {Köhler}, \citenamefont {Zarini},
  \citenamefont {Hollatz}, \citenamefont {Schinkel}, \citenamefont {D’Arcy},
  \citenamefont {Schwinkendorf}, \citenamefont {Osterhoff}, \citenamefont
  {Irman}, \citenamefont {Schramm},\ and\ \citenamefont
  {Karsch}}]{Kurz_2018_Lanex}%
  \BibitemOpen
  \bibfield  {author} {\bibinfo {author} {\bibfnamefont {T.}~\bibnamefont
  {Kurz}}, \bibinfo {author} {\bibfnamefont {J.~P.}\ \bibnamefont {Couperus}},
  \bibinfo {author} {\bibfnamefont {J.~M.}\ \bibnamefont {Krämer}}, \bibinfo
  {author} {\bibfnamefont {H.}~\bibnamefont {Ding}}, \bibinfo {author}
  {\bibfnamefont {S.}~\bibnamefont {Kuschel}}, \bibinfo {author} {\bibfnamefont
  {A.}~\bibnamefont {Köhler}}, \bibinfo {author} {\bibfnamefont
  {O.}~\bibnamefont {Zarini}}, \bibinfo {author} {\bibfnamefont
  {D.}~\bibnamefont {Hollatz}}, \bibinfo {author} {\bibfnamefont
  {D.}~\bibnamefont {Schinkel}}, \bibinfo {author} {\bibfnamefont
  {R.}~\bibnamefont {D’Arcy}}, \bibinfo {author} {\bibfnamefont {J.-P.}\
  \bibnamefont {Schwinkendorf}}, \bibinfo {author} {\bibfnamefont
  {J.}~\bibnamefont {Osterhoff}}, \bibinfo {author} {\bibfnamefont
  {A.}~\bibnamefont {Irman}}, \bibinfo {author} {\bibfnamefont
  {U.}~\bibnamefont {Schramm}}, \ and\ \bibinfo {author} {\bibfnamefont
  {S.}~\bibnamefont {Karsch}},\ }\href {\doibase 10.1063/1.5041755} {\bibfield
  {journal} {\bibinfo  {journal} {Review of Scientific Instruments}\ }\textbf
  {\bibinfo {volume} {89}},\ \bibinfo {pages} {093303} (\bibinfo {year}
  {2018})}\BibitemShut {NoStop}%
\bibitem [{\citenamefont {Lehe}\ \emph {et~al.}()\citenamefont {Lehe},
  \citenamefont {Kirchen}, \citenamefont {Andriyash}, \citenamefont {Godfrey},\
  and\ \citenamefont {Vay}}]{lehe_spectral_2016}%
  \BibitemOpen
  \bibfield  {author} {\bibinfo {author} {\bibfnamefont {R.}~\bibnamefont
  {Lehe}}, \bibinfo {author} {\bibfnamefont {M.}~\bibnamefont {Kirchen}},
  \bibinfo {author} {\bibfnamefont {I.~A.}\ \bibnamefont {Andriyash}}, \bibinfo
  {author} {\bibfnamefont {B.~B.}\ \bibnamefont {Godfrey}}, \ and\ \bibinfo
  {author} {\bibfnamefont {J.-L.}\ \bibnamefont {Vay}},\ }\href {\doibase
  10.1016/j.cpc.2016.02.007} {\ \textbf {\bibinfo {volume} {203}},\ \bibinfo
  {pages} {66}}\BibitemShut {NoStop}%
\end{thebibliography}%

\end{document}